\documentclass[iop]{emulateapj}

\def\spose#1{\hbox to 0pt{#1\hss}}
\newcommand\lsim{\mathrel{\spose{\lower 3.0pt\hbox{$\mathchar"218$}}
     \raise 2.0pt\hbox{$\mathchar"13C$}}}
\newcommand\gsim{\mathrel{\spose{\lower 3.0pt\hbox{$\mathchar"218$}}
     \raise 2.0pt\hbox{$\mathchar"13E$}}}
\newcommand\msun{\,M_\odot}

\shorttitle{Nuclear Star-Forming Ring}
\shortauthors{Kim et al.}
\slugcomment{Accepted for publication in the Astrophysical Journal Letters,
vol. 734, 2011}

\begin{document}

\title{Nuclear Star-Forming Ring of the Milky Way: Simulations}

\author{Sungsoo S. Kim,$^{1,2,3,}$\altaffilmark{8}
Takayuki R. Saitoh,$^4$
Myoungwon Jeon,$^{1,5}$
Donald F. Figer,$^3$
David Merritt,$^6$
\\
and Keiichi Wada$^7$}

\affil{$^1$Dept. of Astronomy \& Space Science, Kyung Hee University,
Yongin, Kyungki 446-701, Korea \\
$^2$School of Space Research (WCU program), Kyung Hee University,
Yongin, Kyungki 446-701, Korea \\
$^3$Center for Detectors, Rochester Institute of Technology,
Rochester, NY 14623, USA \\
$^4$Center for Computational Astrophysics, National Astronomical
Observatory of Japan, Mitaka, Tokyo 181-8588, Japan \\
$^5$Dept. of Astronomy, University of Texas, Austin, TX 78712, USA \\
$^6$Dept. of Physics and Center for Computational Relativity
and Gravitation, Rochester Institute of Technology, Rochester, NY 14623, USA \\
$^7$Graduate School of Science and Engineering, Kagoshima
University, Kagoshima 890-8580, Japan}
\altaffiltext{8}{A part of S. S. K.'s work was conducted at Rochester
Institute of Technology during his sabbatical leave from Kyung Hee University.}

\begin{abstract}
We present hydrodynamic simulations of gas clouds in the central kpc region
of the Milky Way that is modeled with a three-dimensional bar potential.
Our simulations consider realistic gas cooling and heating, star formation,
and supernova feedback.  A ring of dense gas clouds forms as a result of
$X_1$--$X_2$ orbit transfer, and our potential model results in a ring radius
of $\sim 200$~pc, which coincides with the extraordinary reservoir of dense
molecular clouds in the inner bulge, the Central Molecular Zone (CMZ).
The gas clouds accumulated in the CMZ can reach high enough densities to
form stars, and with an appropriate choice of simulation parameters, we
successfully reproduce the observed gas mass and the star formation rate (SFR)
in the CMZ, $\sim 2 \times 10^7 \msun$ and $\sim 0.1 \msun \, {\rm yr}^{-1}$.
Star formation in our simulations takes place mostly in the outermost
$X_2$ orbits, and the SFR per unit surface area outside
the CMZ is much lower.  These facts suggest that the inner Galactic bulge
may harbor a mild version of the nuclear star-forming rings seen in some
external disk galaxies.  Furthermore, from the relatively small size of
the Milky Way's nuclear bulge, which is thought to be a result of sustained
star formation in the CMZ, we infer that the Galactic inner bulge probably
had a shallower density profile or stronger bar elongation in the past.
\end{abstract}


\section{Introduction}
\label{sec:intro}

CO emission surveys along the Galactic plane reveal that molecular gas
is abundant in the plane down to a Galactocentric radius $r$ of
$\sim 3$~kpc (see e.g., Dame, Hartmann, \& Thaddeus 2001).  Inside
this radius, the amount of molecular emission greatly reduces, as
do star formation activities.  However, a significant amount of
molecular gas, as well as various evidences of recent star formation,
appears again inside $r$ of $\sim 200$~pc (Morris \& Serabyn 1996,
hereafter MS96; Yusef-Zadeh et al. 2009, hereafter YZ09).  This region
of abundant molecular gas is called the Central Molecular Zone (CMZ).
The total gas mass inside $r=250$~pc is estimated to be
$1.7$--$5 \times 10^7 \msun$ (Dahmen et al. 1998; Pierce-Price et al. 2000;
Launhardt, Zylka, \& Mezger 2002).

The gas in the CMZ must have migrated inward from the Galactic disk through
interactions with other clouds, field stars, and magnetic fields (MS96).
A mixed atomic/molecular layer inside $r \simeq 3$~kpc turns into a
mostly molecular, high-density medium at the CMZ (Liszt \& Burton 1978; MS96).
Binney et al. (1991) hypothesize that this abrupt change in gas content and
density is caused by a transition between so-called $X_1$ and $X_2$ orbits
near $r = 200$~pc.  $X_1$ is a family of stable, closed orbits elongated
along the bar's major axis, whereas $X_2$ orbits are elongated along the
bar's minor axis, much deeper in the potential (Contopoulos \& Papayannopoulos
1980).  The spiraling-in gas clouds would generally follow the $X_1$ orbits
until the orbits become sharply cusped and even self-intersecting at the
innermost $X_1$ region.  Gas clouds experience compression and shocks near
the cusps of these orbits, lose angular momentum, and plunge to the $X_2$
orbits inside.  The infalling gas clouds can additionally lose angular
momentum when they collide with those already on the $X_2$ orbits.
This compression and subsequent cooling will transform the
gas into molecular form, and the molecular clouds accumulated in the $X_2$
region would correspond to the CMZ.

The abundance and high densities of molecular gas in the CMZ naturally
lead to star formation.  Indeed the CMZ harbors two extraordinary, young
(few Myr), massive (a few $10^4 \msun$) star clusters, the Arches and the
Quintuplet (Figer et al. 1999; Kim et al. 2006, among others) as well as
a population of young stellar objects (YZ09).  Serabyn \& Morris (1996)
propose that star formation has been occurring in the CMZ throughout the
lifetime of the Galaxy, and the resulting stellar population is evident
as the central $r^{-2}$ cluster or the nuclear bulge, a prominent cusp
in the stellar density profile, whose extent is similar to that of the CMZ.

There have been a few numerical studies on the hydrodynamics of the
gas layer in the Galactic inner bulge.  Jenkins \& Binney (1994)
examined the hypothesis of Binney et al. (1991) with two-dimensional
(2--D) sticky particle simulations and confirmed that the transition of
gas motion from $X_1$ to $X_2$ orbits indeed takes place in a bar potential.
Lee et al. (1999) and Englmaier \& Gerhard (1999) were able to observe
the same phenomenon in their 2--D smoothed particle hydrodynamics (SPH)
simulations.  Rodriguez-Fernandez \& Combes (2008) performed 2--D
hydrodynamic simulations (particle motions followed in 3--D) with a more
sophisticated potential model for the inner Galaxy in an attempt to
reproduce the observed velocity structure of the CMZ.  All these studies
primarily concentrated on the behavior of the gas flows, and were not
able to follow the density and temperature evolution of the gas clouds
and subsequent star formation,\footnote{A recent Milky Way simulation by
Baba, Saitoh, \& Wada (2010), whose main interest was gas kinematics,
does follow thermal evolution and star formation of the gas.} due to the
limits in their numerical schemes and/or computing power at those times.

In the present Letter, we present 3--D hydrodynamic simulations for the
motions of gas clouds and star formation in the inner bulge of the Galaxy.
Our simulations consider self-gravity, gas cooling and heating,
and supernova (SN) feedback.  We show that a bar potential can indeed compress
and cool the gas sufficiently enough to form stars and that the obtained
star formation rates (SFRs) are consistent with observations.

\section{Methods and Models}

We use a parallel $N$-body/SPH code ASURA (Saitoh et al. 2008, 2009)
for our simulations.
We use an opening angle of $\theta =0.5$ for a cell opening and a softening
length of 3~pc, and the kernel size of an SPH particle is determined by
imposing the number of neighbors to be $32 \pm 2$.
We implement a cooling function by Spaans \& Norman (1997)
for gas with a solar metallicity for a temperature range of 10--$10^8$~K,
and uniform heating from far-ultraviolet (FUV) radiation is considered.
The standard radiation strength used in our simulations is
$100 \, G_0$,\footnote{$G_0 \equiv 1.6 \times 10^{-3} {\rm erg \, cm^{-2}
\, s^{-1}}$ is the Habing field (Habing 1968).}
which is $\sim 60$ times the solar neighborhood value (the SFR
per unit surface area in the CMZ is $\sim 60$ times larger than
that in the solar neighborhood; YZ09).  The effect of
SN feedback is implemented by supplying thermal energy into the surrounding
32 gas particles.

A star particle is spawned when a gas particle satisfies all of the following
four conditions:
1) hydrogen number density is larger than the threshold density, $n_{\rm th}$,
2) temperature is lower than the threshold temperature, $T_{\rm th}$,
3) the flow is converging ($\mathbf{\nabla} \cdot \mathbf{v} < 0$), and
4) there is no heat supply from nearby SN explosions.
We adopt $n_{\rm th} = 100$~cm$^{-3}$ and $T_{\rm th}=100$~K as our standard
threshold values following Saitoh et al. (2008, 2009), who showed that only
such a high density and low temperature threshold can reproduce the
structural characteristics of galactic disks.  The local SFR
is assumed to be proportional to the local gas density divided by
the local dynamical time,
\begin{equation}
\label{sfr}
	\frac{d \rho_*}{dt} = C_* \frac{\rho_{\rm gas}}{t_{\rm dyn}},
\end{equation}
where $C_*$ is the star formation efficiency (SFE) parameter, and we adopt
$C_*=0.033$ as our standard value following Saitoh et al.  A newly
spawned star particle is set to have one-third of the original gas particle
mass.  Detailed discussion on the spawning of star particles and SN
feedback is given in Saitoh et al.

Following Zhao, Spergel, \& Rich (1994), we adopt an $m=2$ bar with a
power-law density profile for the inner Galactic bulge:
\begin{equation}
\label{density}
        \rho=\rho_0 \left ( \frac{r}{r_0} \right )^{-\alpha}
                    \left [ 1 - Y(\theta,\phi) \right ],
\end{equation}
where
\begin{equation}
\label{Y}
       Y(\theta,\phi)= -b_{20} P_{20}(\cos \theta)
                       + b_{22} P_{22}(\cos \theta) \cos 2 \phi,
\end{equation}
and $P$ is the associated Legendre function.  These are linear combinations of
spherical harmonic functions of the $l=2$, $m=0$,2 modes.  $b_{20}$ determines
the degree of oblateness/prolateness while $b_{22}$ determines the degree of
non-axisymmetry.  Our standard simulation (simulation 1) has the following
parameters:
$\alpha=1.75$, $b_{20}=0.3$, $b_{22}=0.1$, $\rho_0=40 \msun$~pc$^{-3}$,
$r_0=100$~pc, and a bar pattern speed of $63\,{\rm km/s/kpc}$.
This set of parameters gives a bar axis ratio of 1:0.74:0.65 for the
isodensity surface that intersects points [$x=0$, $y=\pm 200$~pc, $z=0$],
an $X_1$--$X_2$ transition at $r \sim 200$~pc, inner Lindblad resonance
at 660~pc, and a corotation radius of 2.8~kpc.  Enclosed masses inside
200~pc and 1000~pc are $10^9 \msun$ and $7 \times 10^9 \msun$, respectively.

Our simulations initially have gas particles only, which are distributed
on $X_1$ orbits whose semi-minor axes range from 300 to 1200~pc.
The number of particles on each orbit is proportional to the length of
the orbit, and on a given orbit, particles are distributed over the same
time-interval.  The initial vertical distribution of the gas particles
follows a Gaussian function with a scale height of 40~pc.  Our standard
simulation has $2 \times 10^5$ gas particles and a total gas mass of
$10^8 \msun$, thus each gas particle initially has a mass of $500 \msun$
(the latter is true for all of simulations, thus star particles in
our simulations typically have $\sim 170 \msun$).  The initial velocities
in the plane follow the motions on the $X_1$ orbits, while the vertical
component is initially set to zero.  We evolve such initial setup without
cooling, heating, star formation, and SN feedback for the first 50~Myr
to obtain a hydrodynamically relaxed particle distribution.

\begin{deluxetable}{cclrcc}
\tablecolumns{6}
\tablewidth{0pt}
\tablecaption{\label{table:sim}Simulation Parameters and Results}
\tablehead{
\colhead{} &
\colhead{$M_{g,i}$} &
\colhead{} &
\colhead{$I_{\rm FUV}$} &
\colhead{$M_{g,X2}$} &
\colhead{SFR$_{X2}$} \\
\colhead{Sim} &
\colhead{($\rm 10^7 \msun$)} &
\colhead{$C_*$} &
\colhead{($G_0$)} &
\colhead{($\rm 10^7 \msun$)} &
\colhead{($\rm M_\odot \, yr^{-1}$)}
}
\startdata
1 & 10 & 0.033  &  100 & 1.33 & 0.060 \\
2 & 20 & 0.033  &  100 & 2.03 & 0.116 \\
3 & 10 & 0.0033 &  100 & 1.85 & 0.045 \\
4 & 10 & 0.033  & 1000 & 1.11 & 0.057
\enddata
\tablecomments{$M_{g,i}$ is the initial total gas mass, $I_{\rm FUV}$ is
the FUV radiation strength, and $M_{g,X2}$ and SFR$_{X2}$ are the
gas mass and the star formation rate in the $X_2$ region averaged
for $T=200$--600~Myr.  The initial number of gas particles is 200,000
for simulations 1, 3, \& 4, and 400,000 for simulation 2.}
\end{deluxetable}

\section{Results}
\label{sec:results}

\begin{figure*}
\center\epsscale{1.15}
\plotone{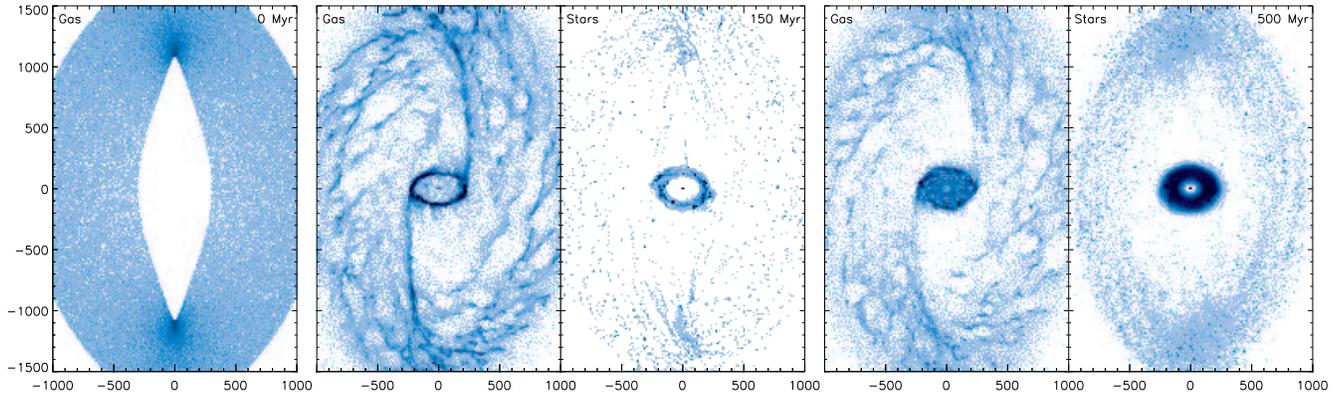}
\caption{\label{fig:snap01}Surface density maps of gas and (accumulated)
star particles in simulation 1 at $T=0$, 150, and 500~Myr.  Length units
are in parsecs, and the gray scale represents densities linearly from 3 to
$300 \msun \, {\rm pc^{-2}}$.  The major axis of the bar potential is
aligned along the $y$-axis.}
\end{figure*}

\begin{figure*}
\center\epsscale{1.15}
\plotone{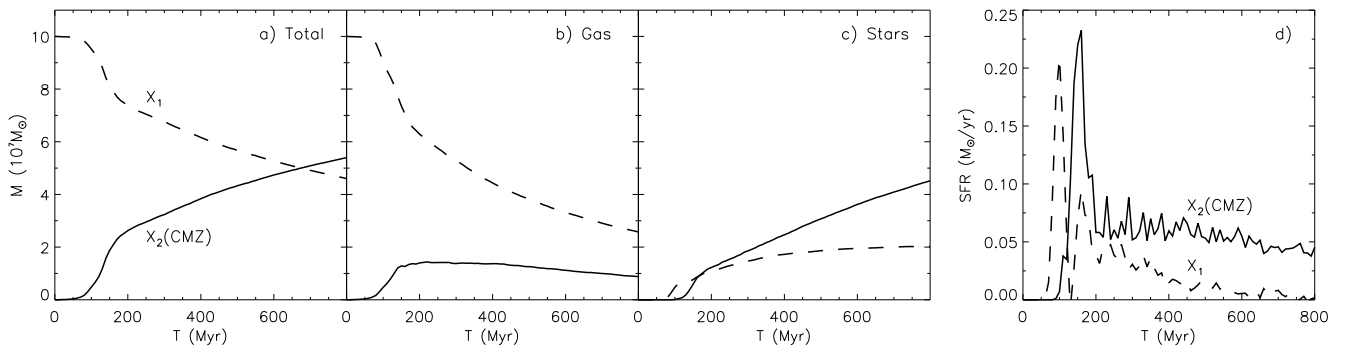}
\caption{\label{fig:mass01}Evolution of total (gas and stars), gas, and
stellar masses (panels {\it a} through {\it c}), and the star formation rate
(panel {\it d}) in the $X_1$ (dashed lines) and $X_2$ (solid lines) regions
of simulation 1.  After the initial relaxation period ($T>200$~Myr),
most of star formation takes place in the $X_2$ region (CMZ), and the
obtained SFR in the CMZ is consistent with the observed values by YZ09,
0.01--$0.1 \msun$~yr$^{-1}$.}
\end{figure*}

Three snapshots of our standard simulation are presented in Figure
\ref{fig:snap01} for $T=0$, 150 and 500~Myr.  The sense of galactic rotation
in the snapshots is clockwise.
The middle panel ($T=150$~Myr) shows that gas particles near
[$x=0$, $y=\pm 1250$~pc] quickly undergo shocks and compressed clouds plunge
to the $X_2$ orbits along the ``dust lanes'' in the leading inner edges of
the $X_1$ orbits.  The gas accumulated on the $X_2$ orbits, i.e., the CMZ,
initially forms an elongated ring of clouds at $r \sim 200$~pc.
The masses of gas clumps and streams in the CMZ of our simulations,
$10^5$--$10^6 \msun$, agree with those estimated for some clumps from
radio observations (e.g., Stark et al. 1991).  Gas density is the highest
in this ring, and this is where star formation is the most active.

The right panel ($T=500$~Myr) shows that as the gas infall and star
formation in the nuclear ring continue, the radial distribution of gas and
stars in the $X_2$ region broadens (mostly inward\footnote{The gas
particles in the CMZ lose orbital energy via shear viscosity from
differential rotation and gradually migrate inward.  Some particles
should gain angular momentum and the CMZ should broadens outward as well,
but those particles would merge with the part of the dust lane stream that
returns back to the $X_1$ region.}).  However, star formation
still takes place mostly near $r \sim 200$~pc, because newly
infalling gas from the dust lanes collides the existing nuclear ring of
gas at its outer rim.
We note that only $\sim 15~\%$ of the gas flowing down the dust lanes
directly enters the CMZ, while the rest flows back to the $X_1$ region.
This fraction is consistent with that found in Regan, Vogel, \& Teuben (1997;
10--25~\%).

Evolutions of gas, stellar, and total masses on $X_1$ and $X_2$ orbits
are shown in Figure \ref{fig:mass01}.
The mass transfer from $X_1$ to $X_2$ regions takes place quite rapidly
between $T \simeq 100$ and $\simeq 200$~Myr and then goes into a slower,
stable phase (panel {\it a}).  While the total mass in the
$X_2$ region grows nearly constantly after $T=200$~Myr, the gas mass
in the $X_2$ region, $M_{g,X2}$, is held at a nearly constant value and
then decreases very slowly (panel {\it b}).
This implies that star formation in the $X_2$ region
becomes so efficient that all of the newly supplied gas turns into stars
once the gas mass in the $X_2$ region reaches a certain critical value
($\sim 1.5 \times 10^7 \msun$) at $\simeq 200$~Myr.

Both the $X_1$ and $X_2$ regions initially ($T<200$~Myr) have high SFRs (see
panel {\it d}), but this is because when the star formation takes place for
the first time ($T=50$~Myr for the $X_1$ region, $\simeq 100$~Myr for the
$X_2$ region), there is no or not enough SN feedback that regulates star
formation.  Thus this initial relaxation period should be disregarded
if one assumes a continuous gas supply from the outer region (an intermittent
gas supply may result in a repetition of a brief, high-SFR phase followed by
a longer, lower-SFR phase).  While the SFR in the $X_1$ region continuously
decreases as the amount of gas there becomes depleted, the SFR in the $X_2$
region becomes quite stationary after the initial relaxation period, and
ranges between 0.04 and $0.09 \msun \, {\rm yr}^{-1}$.  This range is
in excellent agreement with recent SFRs estimated for the CMZ from
mid-infrared observations by YZ09, 0.01--$0.1 \msun$~yr$^{-1}$.

Simulation 1 clearly shows that 1) under a bar potential modeled for the
inner bulge of the Milky Way, gas inflow along the Galactic plane leads
to the formation of a gas ring, whose radial extent is consistent
with the CMZ, and 2) the gas ring has high enough densities to form stars
at a rate consistent with observed values.
The parameters of our potential model ($\rho_0$, $r_0$, $\alpha$, $b_{20}$,
and $b_{22}$) were chosen such that the transition between $X_1$ and $X_2$
orbits appears at $\sim 200$~pc, therefore the extent of the gas ring in our
simulation is a result of the potential parameters that we chose.

On the other hand, it is reasonable to expect that the SFR in the CMZ
is determined by the gas influx rate from the $X_1$ to $X_2$ regions, and
that this influx rate is dependent on the initial gas surface density
(or initial gas mass) in the $X_1$ region.  The fact that the SFR from
simulation 1 is consistent with the observed values implies that the
adopted initial gas mass in our $X_1$ region (major axes from 300 to
1200~pc) properly reproduces the actual gas influx rate from the $X_1$ to
$X_2$ regions in the GC.

However, the average $M_{g,X2}$ after relaxation ($200 < T < 600$~Myr) in
simulation 1, $1.3 \times 10^7 \msun$, is 20~\% to four times smaller than
the estimated $M_{g,X2}$ values from various observations, $1.7$--$5 \times
10^7 \msun$ (the largest mass estimate is derived from dust emission and
is sensitive to the assumed dust temperature in the GC, while the smallest
estimate is obtained from C$^{18}$O molecular line emission;
see Section \ref{sec:intro} for references).

To see the dependence of $M_{g,X2}$ on initial $M_{g,X1}$, we performed
simulation 2, which starts with a twice as large initial $M_{g,X1}$
than simulation 1.  Simulation 2 indeed results in $\sim 2$ times larger
SFRs than simulation 1 as expected, and $\sim 1.5$ times larger
$M_{g,X2}$ (see Table \ref{table:sim}).  Now, not only the SFR but also
$M_{g,X2}$ is consistent with observations.  Note that the CMZ mass is
not as sensitive to the gas influx rate as the SFR is.

Simulation 3, whose SFE parameter ($C_*$) is 10 times smaller than
simulation 1 in consideration of the extreme star formation environment
in the GC such as strong tidal forces/magnetic fields and elevated
gas temperatures, shows another way to increase $M_{g,X2}$---it gives
$\sim 25$~\% smaller SFR, but $\sim 40$~\% larger
$M_{g,X2}$ ($1.9 \times 10^7 \msun$) than simulation 1.  The increase in
$M_{g,X2}$ is due to the fact that the star formation in the CMZ now
requires higher gas densities than in simulation 1 to compensate for the
lowered $C_*$ in Equation (\ref{sfr}).  From our experiments with
simulations 1 through 3, we conclude that one can fit the observed gas
mass and SFR in the CMZ simultaneously by controlling the gas influx rate
from the outer bulge (i.e., initial $M_{g,X1}$ in our simulations) and
the SFE ($C_*$).

The actual SFR in the whole $X_1$ region of the Milky Way is not known yet,
as no systematic observations have been made for the region.
Simulations 1 through 3 show some star formation in the $X_1$ region, but
it is not possible to accurately calculate steady-state SFRs in the $X_1$
region from these simulations, because our simulations do not have
continuous gas supply from the outer bulge.  The total SFRs in the $X_1$
region of simulations 1 through 3 are only fractions of those in the $X_2$
region, implying that the SFR per unit volume is much higher in the $X_2$
region (the volume occupied by the $X_1$ orbits is $\sim 30$ times larger
than that occupied by the $X_2$ orbits in these simulations).

\begin{figure}
\center\epsscale{1.15}
\plotone{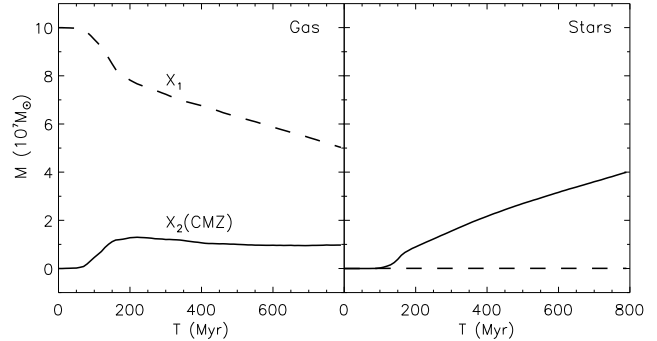}
\caption{\label{fig:mass04}Evolution of gas and stellar masses in the $X_1$
(dashed lines) and $X_2$ (solid lines) regions of simulation 4.
The increased $G_0$ effectively suppresses star formation in the $X_1$
region, where the gas density is lower.}
\end{figure}

In many nuclear starburst galaxies, star formation takes the form of a ring,
i.e., a nuclear star-forming ring.  If the Milky Way is such a case, star
formation in the inner bulge would be confined in the CMZ and there should
be no or very little star formation taking place outside the CMZ.
We find from simulation 4 that one way to achieve this is to impose a
higher FUV radiation field.  Simulation 4, whose $G_0$ is 10 times larger
than simulation 1, gives nearly the same gas mass and SFR in the $X_2$
region as simulation 1, but almost entirely suppresses star formation in
the $X_1$ region\footnote{The suppression of star formation by FUV radiation
is much more effective in less dense gas.} (see Figure \ref{fig:mass04}).
It is not certain whether the radiation strength in the inner bulge can
actually reach as high as $1000 \, G_0$, but other physical mechanisms that
can quench star formation in the GC, but are not considered in our
simulations, such as strong tidal forces and magnetic fields, may also
contribute to the suppression of star formation in the $X_1$ region
as seen in simulation 4.

Our star formation and SN feedback models successfully reproduce the
Kennicutt-Schmidt (K-S) relation.  At $T=400$~Myr, the CMZ in simulation 1
has a gas surface density of $100 \msun /{\rm pc}^2$ and a surface SFR of
$0.45 \msun \, {\rm yr}^{-1} \, {\rm kpc}^2$, the latter of which is 0.45 dex
higher than the mean K-S relation, whereas the CMZ in simulation 3 has a
surface SFR that is 0.12 dex higher than the mean K-S relation.  Considering a
rather large scatter in observed SFRs around a mean K-S relation (rms scatter
$\sim 0.3$ dex; Kennicutt 1998), the SFRs from our simulations are consistent
with the K-S relation.

\section{The CMZ and the Nuclear Bulge}

If the star formation in the inner bulge of the Milky Way is indeed confined
in the CMZ, and if it has been sustained for the lifetime of the Galaxy
as suggested by Figer et al. (2004), the resulting stellar population should
manifest in a certain way.  Indeed,
the infrared intensity profiles in the GC have revealed a prominent, central
stellar cusp with an outer extent of about 0.7 deg ($\sim 100$~pc) on top
of the shallower profile made by the Galactic bulge (Serabyn \& Morris 1996,
and references therein).
Serabyn \& Morris propose that this `nuclear bulge' is the resulting stellar
population from the sustained star formation in the CMZ.\footnote{Another
way of creating the nuclear bulge is the inspiral of massive star clusters via
dynamical friction (Bekki 2010, and references therein).}

The extent of the nuclear bulge is somewhat smaller than that of the CMZ,
and this may indicate that the radius of the CMZ was smaller in the past.
From our further simulations with various density slopes ($\alpha$) and
bar elogations ($b_{22}$), we find that a smaller CMZ radius can be obtained
when the density profile is shallower ($\alpha \lsim 1.6$; the same
amount of orbital energy loss at the cusps of $X_1$ orbits will plunge the gas
further inside), or when the bar is more elongated ($b_{22} \gsim 0.14$;
the gas at the cusps of $X_1$ orbits will lose larger orbital energy).
Therefore, the relatively smaller size of the nuclear bulge may be an
important clue on the secular evolution of the Galactic center.

Some of the accumulated gas in the CMZ will keep migrating inward and
reach the CircumNuclear Disk (CND) of molecular clouds at $r=2$--5~pc,
and a part of it will sink further to form stars in the central parsec
or feed the central supermassive black hole.  One possible way to transfer
gas from the CMZ to the CND would be a `nested bar,' as examined by Namekata
et al. (2009).

\acknowledgments
This work was supported by the Korea Research Foundation (KRF-2008-013-C00037)
and by the WCU program (R31-1001) through the NRF funded by the MEST of Korea.
We thank the anonymous referee for their valuable comments that greatly
improved our Letter.
The material in this work is partly supported by NASA under awards NNG 05-GC37G
and NNX 10-AF84G, through the Long--Term Space Astrophysics program, and by
HPCI Strategic Program Field 5, "The origin of matter and the universe".
A part of this research was performed in the Rochester Imaging Detector
Laboratory with support from a NYSTAR Faculty Development Program grant.


\end{document}